\documentclass[sigconf]{acmart}

\usepackage{xcolor}
\AtBeginDocument{%
  }

\newcommand{\laddF}[1]{#1}
\newcommand{\lremF}[1]{}

\newcommand{\ladd}[1]{#1}
\newcommand{\lrem}[1]{}

\copyrightyear{2024}
\acmYear{2024}
\setcopyright{rightsretained}
\acmConference[CHI '24]{Proceedings of the CHI Conference on Human Factors in Computing Systems}{May 11--16, 2024}{Honolulu, HI, USA}
\acmBooktitle{Proceedings of the CHI Conference on Human Factors in Computing Systems (CHI '24), May 11--16, 2024, Honolulu, HI, USA}
\acmDOI{10.1145/3613904.3642245}
\acmISBN{979-8-4007-0330-0/24/05}

\begin{document}

\title{HIV Client Perspectives on Digital Health in Malawi}

\author{Lisa Orii}
\email{lisaorii@cs.washington.edu}
\orcid{0000-0002-6490-4776}
\affiliation{%
  \institution{University of Washington}
  \city{Seattle}
  \state{Washington}
  \country{USA}
}

\author{Caryl Feldacker}
\email{cfeld@uw.edu}
\orcid{0000-0002-8152-6754}
\affiliation{%
  \institution{University of Washington}
  \city{Seattle}
  \state{Washington}
  \country{USA}
}
\author{Jacqueline Madalitso Huwa}
\email{jhuwa@lighthouse.org.mw }
\orcid{0000-0001-8786-2487}
\affiliation{%
  \institution{Lighthouse Trust}
  \city{Lilongwe}
  \country{Malawi}
}
\author{Agness Thawani}
\email{athawani@lighthouse.org.mw }
\orcid{0009-0007-1221-0901}
\affiliation{%
  \institution{Lighthouse Trust}
  \city{Lilongwe}
  \country{Malawi}
}
\author{Evelyn Viola}
\email{eviola@lighthouse.org.mw }
\orcid{0000-0003-1283-6182}
\affiliation{%
  \institution{Lighthouse Trust}
  \city{Lilongwe}
  \country{Malawi}
}
\author{Christine Kiruthu-Kamamia}
\email{ckamamia@lighthouse.org.mw }
\orcid{0009-0008-7682-2496}
\affiliation{%
  \institution{Lighthouse Trust}
  \city{Lilongwe}
  \country{Malawi}
}
\author{Odala Sande}
\email{o_sande@lighthouse.org.mw}
\orcid{0000-0003-1481-7161}
\affiliation{%
  \institution{Lighthouse Trust}
  \city{Lilongwe}
  \country{Malawi}
}

\author{Hannock Tweya}
\email{tweyah@uw.edu}
\orcid{0000-0002-8247-1273}
\affiliation{%
  \institution{University of Washington}
  \city{Seattle}
  \state{Washington}
  \country{USA}
}
\affiliation{
  \institution{International Training and Education Center for Health (I-TECH)}
  \city{Lilongwe}
  \country{Malawi}
}

\author{Richard Anderson}
\email{anderson@cs.washington.edu}
\orcid{0000-0002-7283-7219}
\affiliation{%
  \institution{University of Washington}
  \city{Seattle}
  \state{Washington}
  \country{USA}
}
\renewcommand{\shortauthors}{Orii et al.}

\begin{abstract}
eHealth has strong potential to advance HIV care in low- and middle-income countries. Given the sensitivity of HIV-related information \ladd{and the risks associated with unintended HIV status disclosure}, \lrem{it is important to examine }clients’ privacy perceptions towards eHealth applications \ladd{should be examined} to develop client-centered technologies. Through focus group discussions with antiretroviral therapy (ART) clients from Lighthouse Trust, Malawi’s public HIV care program, we explored perceptions of data security and privacy, including their understanding of data flow\lrem{ from collection through aggregation} and their concerns about data confidentiality across several layers of data use. Our findings highlight the broad privacy concerns that affect ART clients’ day-to-day choices, clients’ trust in \lrem{the}\ladd{Malawi's} health system, and their acceptance of, and familiarity with,\lrem{the} point-of-care technologies used in HIV care. Based on our findings, we provide \lrem{suggestions}\ladd{recommendations} for building robust digital health systems in low- and middle-income countries with limited resources, nascent privacy regulations, and political will to take action to protect client data.
\end{abstract}




\maketitle

\section{Introduction} 

eHealth\footnote{We refer to ``eHealth” as digital information systems used in the health system, including centrally managed electronic medical record systems, mobile devices for data collection, and national health indicator reporting.} initiatives, including computer- and tablet-based interventions, offer great promise for health systems to enhance human immunodeficiency virus (HIV) care, especially in low- and middle-income countries (LMICs) \ladd{where HIV prevalance is high} ~\cite{kemp2018implementation}. eHealth is considered a key tool for addressing challenges across the HIV care cascade, including testing, diagnosis, adherence to antiretroviral therapy (ART), and viral suppression ~\cite{bervell2019comparative, tweya2016developing, wessels2007improving, castelnuovo2012implementation}. Especially in HIV care, security and privacy\footnote{We use the term ``security” to refer to protection mechanisms against threats to client data, and ``privacy” to refer to clients’ right to control how their data is viewed, shared, and used.} issues are of utmost importance, as eHealth applications, such as electronic medical record (EMR) systems, contain sensitive information and data exposure could leave people living with HIV (PLHIV) to the face of stigma and discrimination. While the urgency to address digital security concerns has garnered attention from researchers \ladd{within HCI ~\cite{10.1145/2441776.2441837, 10.1145/2531602.2531643, 10.1145/2675133.2675204, 10.1145/2675133.2675252, 10.1145/3411764.3445420, 10.1145/3491102.3501864, 10.1145/2702123.2702392} and beyond ~\cite{GARDIYAWASAMPUSSEWALAGE20161161, parks2011healthcare, parks2011understanding, GHAZVINI2013212}}\lrem{, including those in HCI }\lrem{\mbox{~\cite{10.1145/2441776.2441837, 10.1145/2531602.2531643, 10.1145/2675133.2675204, 10.1145/2675133.2675252, GARDIYAWASAMPUSSEWALAGE20161161, parks2011healthcare, parks2011understanding, GHAZVINI2013212, 10.1145/3411764.3445420, 10.1145/3491102.3501864, 10.1145/2702123.2702392}}}, the privacy considerations of PLHIV are broad – extending beyond threats in the digital space to include concerns around disclosing positive HIV status and the health system’s efforts to protect client confidentiality\lrem{ overall}. As adoption of eHealth systems gains momentum, consideration of client preferences, perspectives, and experiences, including privacy concerns, should be prioritized ~\cite{SHEN20191} for making operational decisions in deploying client-centered technology and easing the adoption of eHealth interventions ~\cite{leblanc2020patient, 10.1093/jamia/ocv014}. However, consideration of, and amplification for, client perspectives is still lacking, including in LMICs where eHealth implementation is expanding rapidly ~\cite{fritz2015need}.

To give voice to clients’ perspectives in an LMIC context, we investigated how HIV positive clients on ART at one of Malawi’s largest public service delivery partners for HIV care, Lighthouse Trust (hereafter referred to as Lighthouse), perceive data security and privacy in the context of electronic and mobile data collection. By exploring clients’ views, we expand the understanding of how PLHIV navigate privacy, create opportunities for enhancing client-centered eHealth care, and provide a forum for clients to express their recommendations for data privacy improvements in the health system. The motivation for this study is an effort by Lighthouse to test expanded use of its EMRs for client management with optimized tablet-based data collection in rural and low connectivity settings and to strengthen security processes for HIV care. Lighthouse is a public ART clinic that operates under the Malawi Ministry of Health (MoH) and serves as a testing ground for client care innovations. As such, Lighthouse is an ideal setting to inform privacy improvements within the broader MoH eHealth system.

We address the following research questions in this work:

\begin{itemize}
\item \textbf{RQ1}: How do ART clients receiving HIV care in Malawi \lrem{navigate}\ladd{consider} privacy\lrem{ in the health system and beyond}?
\item \textbf{RQ2}: How do ART clients perceive the efforts of Malawi’s health system to protect client privacy?
\item \textbf{RQ3}: How do ART clients understand the processes involved in eHealth initiatives at Malawi’s HIV care program, including digital data collection tools and data management?
\end{itemize}

We conducted focus group discussions\lremF{ (FGDs)} with ART clients who receive routine care from Lighthouse. Our findings highlight clients’ broad concerns with privacy in the health system, their trust in Lighthouse and Malawi’s MoH to protect privacy, and their understanding of the data pipeline and digital health system. With respect to \textbf{RQ1}, PLHIV have a broad privacy framework that \lrem{extends beyond digital security concerns }\ladd{centers around concerns and fears associated with unintended disclosure of positive HIV status, although these concerns carried less weight when disclosing within the HIV community.}\lrem{to include disclosure considerations and the health system’s privacy protection efforts.} With respect to \textbf{RQ2}, clients’ trust in Lighthouse’s \lrem{digital}\ladd{devices for} data collection\lrem{ tools} and their \lrem{participation in data contribution}\ladd{interests in contributing their data} are an outgrowth of their \lrem{broader }trust in the \ladd{broader} healthcare that they receive. With respect to \textbf{RQ3}, clients’ mental models of Lighthouse’s point-of-care technologies indicate their familiarity with data flow and \laddF{recognition of the} value of \lrem{eHealth}\ladd{devices for data collection and management}, though \ladd{there was} uncertainty \lrem{around specific technology-based protections remains}\ladd{about the technologies' security mechanisms and exact usage practices}.

More broadly, we illuminate ART clients’ perceptions of Lighthouse’s eHealth system and protection of client privacy, contributing recommendations and considerations that could inform both Malawi’s MoH and other LMIC policies and practices as digital health systems are strengthened and scaled.  

\section{Background}

Sub-Saharan Africa has approximately 25.5 million PLHIV ~\cite{AIDSinfo}. Given Sub-Saharan Africa’s high burden of HIV and increasing technology use, eHealth strategies have been employed to support HIV service delivery ~\cite{OLUOCH2012e83}, including monitoring adherence to ART, managing patient data, and managing the ART supply chain ~\cite{fraser2005implementing, Fraser1142, bervell2019comparative, tweya2016developing, wessels2007improving, castelnuovo2012implementation}.

\begin{figure*}[hbt]
  \centering
  \includegraphics[width=0.8\linewidth]{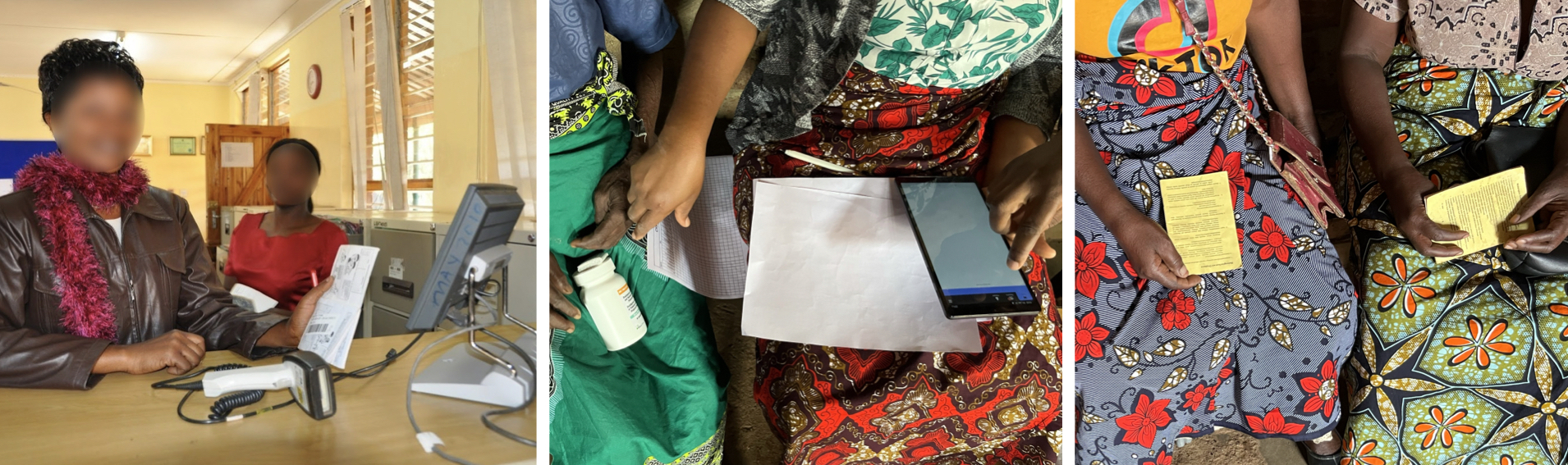}
\caption{A computer at the facility used as an EMR access point at the reception desk by Lighthouse receptionists (left), a tablet used during a client's CBC appointment (middle), and health passport booklets that ART clients bring to appointments (right).}
\Description{Three images of data collection tools used at Lighthouse. The left image is of a computer used at Lighthouse's facility and two receptionists. One receptionist is holding a paper with a barcode on it. A barcode scanner is on the table where the computer is situated. The middle image is of a tablet that is held by a healthcare provider during a CBC appointment with the client on their left. The right image is of two clients holding their yellow health passport booklets.}
\label{fig:tools}
\end{figure*}

This study was conducted in Lilongwe, the capital city of Malawi, a country in Southern Africa with over 50\% of the population living in poverty ~\cite{worldbankMalawi}. Malawi has over 900,000 PLHIV (approximately 8\% of the adult population) ~\cite{unaids}. Lighthouse is a registered public trust and a WHO recognized center of excellence whose role is to support the Malawi MoH’s HIV response. Lighthouse provides services spanning integrated HIV prevention, testing, treatment, and care across two large urban facilities in Lilongwe district in partnership with Malawi’s MoH. All Lighthouse facilities employ Malawi’s real-time point-of-care EMR system, launched in 2010, to ensure that ART services comply with HIV guidelines and to ease client- and clinic-level monitoring and evaluation ~\cite{waters2010experience, 10.1371/journal.pmed.1000319}. The computers at the facilities are stationed at the reception desk, vital signs station, laboratory,  examination rooms, and pharmacy, to enter and access EMRs (Figure \ref{fig:tools} left).

In 2016, Lighthouse introduced community-based care (CBC), \ladd{a nurse-led community-based ART program} where stable clients registered at facilities can access ART services at one of 120 community-based distribution points within Lilongwe, including churches, schools, outdoor areas, and client homes\lrem{ (Figure \ref{fig:cbc-setting})}. At CBC settings, community health nurses collect client data using an offline tablet-based application that operates in rural and low connectivity environments (Figure \ref{fig:tools} middle). Due to restrictions in connectivity and automatic data synchronization, the current version of the tablet-based application locally stores client data until the tablets are returned to the facilities and the data is entered into the EMR system in the computers. With the increasing demand for CBC, Lighthouse is testing a new offline-first tablet-based application to expand the reach of EMRs to CBC settings for facilitating efficient data collection and making client history available to the providers. CBC clients are familiar with both computer-based data collection at the facilities and tablet-based data collection at the CBC settings. The services that are offered through CBC are the same as those at the facilities, which include, but are not limited to adherence monitoring (pill count), screenings, client side-effects and co-morbidity review, refill of prescriptions, documentation of current health status on clients’ health passport booklets and on the tablet-based application, and issuing of next appointment date.


At both facilities and CBC settings, clients use portable, paper-based health passport booklets which are usually kept by clients (Figure \ref{fig:tools} right). Each ART client is issued an ART barcode that is attached to the health passport. The barcode is scanned at facilities’ computers and entered manually in CBC settings’ tablets to confirm client identity and to facilitate longitudinal data capture. The health passport is used to check in the client for the appointment and for the provider to manually record information relevant to the client’s health.  
\section{Related Work} 

\subsection{\ladd{Fear of Inadvertent} Disclosure of \ladd{Positive} HIV Status}
\lrem{Although HIV status disclosure is associated with health benefits such as }\ladd{Disclosure of positive HIV status is recognized as an important first step for} preventing transmission, ensuring effective treatment, and \lrem{achieving}\ladd{receiving} social support ~\cite{kalichman2003stress, simoni2000hiv, ding2011hiv, do2010psychosocial}\lrem{, the stigma around HIV poses a challenge to the disclosure of positive HIV status and to accessing support ~\cite{10.1145/2516604.2516627}}. \ladd{However, disclosure risks leaving PLHIV vulnerable to financial and social ruin ~\cite{10.1145/3434171}, posing a challenge for PLHIV to disclose to access clinical and social support ~\cite{10.1145/2516604.2516627}.}\lrem{HCI researchers have extensively studied the challenges around disclosure and privacy among PLHIV who use social media }\lrem{\mbox{~\cite{10.1145/3290605.3300922, 10.1145/3274450, 10.1145/3313831.3376150, 10.1145/3434171, 10.1145/3415244, 10.1145/2971485.2971542}} or technologies for supporting disclosure processes and coping with HIV-related stigma }\lrem{\mbox{~\cite{10.1145/3544548.3581033, 10.1145/3359156}}.} PLHIV consider various factors in the disclosure process, including disclosing to whom, for what purpose, and how. Often, PLHIV practice selective disclosure of positive HIV status, in which individuals disclose to some people, such as trusted partners, family members, and healthcare providers, but not to others ~\cite{kalichman2003stress, veinot2009lot, ismail2021barriers}. Some PLHIV \lrem{also }view disclosure as a path to destigmatizing and normalizing HIV ~\cite{10.1145/3290605.3300922} \lremF{or}\laddF{and} \lrem{sharing}\ladd{share} stories of perseverance and hope to motivate other PLHIV to seek support and disclose to their sexual partners ~\cite{10.1145/2516604.2516627}. For others, disclosure helps \lrem{address health needs, such as requiring}\ladd{with securing} a support system for when one falls ill ~\cite{linda2013tell}. \ladd{LMIC-based studies find that HIV stigma acutely influences PLHIV's decisions in their care, including accessing ART services far from their homes to avoid unwanted disclosure} \lrem{Research shows that PLHIV also negotiate their disclosure by accessing healthcare away from their homes }~\cite{linda2013tell, christ2022availability, pellecchia2017we, sande2020patient, zakumumpa2020understanding}\lrem{, agreeing with their families to hide one’s positive HIV status from acquaintances ~\cite{linda2013tell}, and keeping discussions between PLHIV confidential ~\cite{10.1145/3411764.3445313}.} \ladd{and having difficulty with treatment adherence out of fear of being seen consuming pills ~\cite{ekstrand2020behavioral, musumari2013if}.}

\ladd{In the HCI community, PLHIV's concerns of disclosure in data-related activities have been examined through the usage of social media\lrem{HCI researchers have recognized the sensitivity of HIV status disclosure and examined challenges in disclosure and privacy in the context of social media} ~\cite{10.1145/3290605.3300922, 10.1145/3274450, 10.1145/3313831.3376150, 10.1145/3434171, 10.1145/3415244, 10.1145/2971485.2971542} and technologies for supporting disclosure processes and coping with HIV-related stigma ~\cite{10.1145/3544548.3581033, 10.1145/3359156}. The literature highlights PLHIV's values and needs for privacy and confidentiality in their concealment of HIV status ~\cite{10.1145/3544548.3581033, 10.1145/3290605.3300922, 10.1145/3313831.3376150} and fears of unintended disclosure of positive HIV status in privacy-effective environments ~\cite{10.1145/3359156}. \laddF{From a study in Ethiopia, Bezabih et al. ~\cite{10.1145/3544548.3580756} highlight disclosure as a process as opposed to a one-time event, in which positive HIV status is discreetly disclosed through a step-by-step approach of using indirect signals and cues or selective disclosure.} However, much of the HCI literature focuses on disclosure and privacy considerations outside of the healthcare context. Accessing ART services carries a risk of inadvertent disclosure of HIV status due to the abundance of sensitive data that is collected and moved through the health system ~\cite{sweeney2013shifting}. Thus, fear of unintended disclosure is highly relevant to ART clients' privacy framework and the decisions they make within and beyond data-related activities in their care.}

\subsection{Clients’ \lrem{Relation to}\ladd{Trust in} the Health System}
Clients’ perceptions towards the health system \lrem{has}\ladd{have} been largely studied in the context of \lrem{data sharing, in particular }clients’ willingness to share data with the health system. \ladd{Clients want transparency in the data sharing process by knowing who accesses their data and by controlling the sharing of their data ~\cite{schwartz2015patient, caine2013patients, 10.1093/jamia/ocv014, 10.1093/jamia/ocad058}, including determining which data to share with whom ~\cite{schwartz2015patient, caine2013patients}.} Clients are often willing to share their data to inform public health decision-making ~\cite{weitzman2012willingness, kohane2005health}, but concerns around data privacy can contribute to their hesitancy to share data ~\cite{yeager2014factors, EDEN201644}. Such concerns arise from clients’ lack of awareness of the data sharing mechanism, which is induced by the distance between clients and healthcare organizations and clients’ unfamiliarity with the technical mechanisms for electronic data sharing ~\cite{esmaeilzadeh2019impacts}. To address such issues, researchers emphasize the importance of building trust between clients and the health system through clear, transparent, and comprehensible policies around data sharing and privacy ~\cite{EDEN201644, esmaeilzadeh2019impacts}. Client trust is critical for protecting client privacy and for creating buy-in from clients to participate in data sharing ~\cite{weitzman2012willingness, 10.1093/jamia/ocac198}. 

Given that healthcare providers are clients’ closest contact in the health system, building client trust at the provider level could positively influence client trust in the larger health system ~\cite{platt2015public}. Providers in reputable healthcare organizations that develop robust privacy policies play a significant role in building client trust in projects for electronic data sharing ~\cite{esmaeilzadeh2019impacts}. Trust in providers and security in the client-provider relationship also alleviate clients’ concerns with technology security\lrem{, such as that of EMR systems} ~\cite{shield2010gradual}. Clients’ \lrem{perceptions of}\ladd{trust in} their providers simultaneously influence decisions about their own health. Nganga et al. ~\cite{nganga2019patient} found that clients in Kenya followed their providers’ recommendations\lrem{ for vaccination} regardless of whether or not they understood the advice\lrem{, because they trusted that their providers have honest motives to guard client health. However, providers’ authority deterred clients from} \ladd{and avoided} openly asking questions\lrem{ about their health or the providers’ recommendations ~\cite{nganga2019patient}}.

\subsection{Client Perspectives on \lrem{eHealth Interventions}\ladd{Devices at the Point-of-Care}}  

\lrem{Although work that highlights client perspectives on eHealth interventions is limited to begin with, most in the area concentrates on clients’ data sharing preferences in high-income countries (HICs) }\lrem{\mbox{~\cite{caine2013patients, schwartz2015patient, 10.1093/jamia/ocad058, leventhal2015designing, caine2015designing, wetzels2018patient, 10.1093/jamia/ocv014}}}\lrem{. Clients want transparency in the data sharing process by knowing who accesses their EMRs and controlling the sharing of their EMRs }\lrem{\mbox{~\cite{schwartz2015patient, caine2013patients, 10.1093/jamia/ocv014, 10.1093/jamia/ocad058}}}\lrem{, including determining which data to share with whom }\lrem{\mbox{~\cite{schwartz2015patient, caine2013patients}}}\lrem{. }

Clients’ perceptions towards \ladd{devices used at the} point-of-care\lremF{ technologies} \lrem{that sync to EMR systems }have \lrem{also }been extensively studied in \lrem{HICs}\ladd{high-income countries} to understand their impact on client-provider interactions ~\cite{10.1145/1978942.1979438, doi:10.1177/1937586719826055, doi:10.1080/24725838.2018.1456988, shachak2009impact, 10.1093/jamia/ocv178}. \lrem{Computers}\ladd{Clients perceive computers} at the point-of-care \lrem{are perceived }as physical barriers to client-provider engagement ~\cite{10.1145/1978942.1979438, shachak2009impact, 10.1093/jamia/ocv178} \lrem{and}\ladd{that} elevate providers’ authority and medical expertise in front of clients ~\cite{10.1145/1978942.1979438}. However, a gradual installation of computer-based EMR systems in exam-rooms can allow clients and providers to adapt to the interventions\ladd{,}\lrem{ and} help maintain healthy client-provider relationships\ladd{,} and \lrem{the encouragement of }\ladd{encourage} health information sharing ~\cite{shield2010gradual}. 

\ladd{The literature in HCI ~\cite{10.1145/3411764.3445420, 10.1145/2702123.2702258, 10.1145/2998581.2998589, 10.1145/3411764.3445111, 10.1145/3491102.3517575} and beyond ~\cite{hall2014assessing, info:doi/10.2196/mhealth.9671, sadasivam2012development, mccool2022mobile} have noted the potential for successfully adopting mobile devices for health in LMICs. However, few focus on}\lrem{ Client} \ladd{client} perspectives \lrem{on}\ladd{towards} \ladd{devices used at the} point-of-care\lrem{ technologies are unstudied in LMICs}. Most similar to this topic is \lrem{work on}\ladd{the study of} client perceptions towards \lrem{digital tools}\ladd{devices} used for conducting health surveys in LMICs ~\cite{10.1145/1357054.1357175, cheng2011barriers, shirima2007use, mclean2017implementing}. \lrem{Past studies raise device}\ladd{Device} familiarity \lrem{as a consideration that }impacts survey respondents’ device acceptability and hence their willingness to answer survey questions ~\cite{cheng2011barriers, van2013collecting, knipe2014challenges, mercader2017female}. Most notably, a study in Angola in 2011 found that the lack of introduction to handheld computers used for surveys contributed to \ladd{respondents'} apprehension to answer HIV-related questions and to disclose sexual behaviors to interviewers ~\cite{cheng2011barriers}. However, \lrem{a}\ladd{in another} study in Sri Lanka in 2014\ladd{, in which}\lrem{ found that} respondents were familiar with handheld devices because of the wide availability of mobile phones\ladd{, the devices were conducive to attracting interest to the survey}\lrem{ and were not unwilling to share information with interviewers} ~\cite{knipe2014challenges}. Even with the lack of familiarity \ladd{with devices}, concerns can be overcome with careful explanations of device purpose ~\cite{knipe2014challenges}. Maintaining data confidentiality is also important to survey respondents. \lrem{Studies report that respondents perceived computers}\ladd{Computers were perceived} to be more secure than paper for recording and storing personal data because limited technology literacy restrict\lrem{s}\ladd{ed} \lrem{data access}\ladd{access to data and because paper was \lrem{perceived to be }more prone to misplacement or damage} ~\cite{cheng2011barriers, mercader2017female}. \lrem{Respondents also perceived paper to be prone to misplacement or damage }\lrem{\mbox{\cite{cheng2011barriers, mercader2017female}.}}More broadly, security and privacy views can derive from individuals’ experiences with technologies, their health, and varying sensitivities around their health conditions ~\cite{10.1145/2956554, caine2013patients}. Especially in the context of sensitive or stigmatized topics such as HIV, respondents’ acceptability of technology for health data collection is critical for obtaining accurate information ~\cite{cheng2011barriers}. Although perspectives on technologies used for health surveys in LMICs have been thoroughly studied, the gap in literature on client perspectives on point-of-care \lrem{technologies}\ladd{devices} in LMICs warrants further research. 
\section{Methods} 
Our primary objective was to explore and understand the perceptions of Lighthouse's HIV \ladd{positive} clients on ART with regard to data security and privacy, electronic and mobile data collection, \lrem{digital }\ladd{devices for} data collection\lrem{ tools}, and data management and sharing processes. The focus group discussions (FGDs) with ART clients were conducted in Lilongwe, Malawi during September and October 2022. This research received approval from the IRB at \lremF{our university}\laddF{the University of Washington} and the Malawi MoH IRB, the National Health Science Research Committee.

\subsection{Setting}
The study was conducted in urban and peri-urban areas of Lilongwe district. FGDs were conducted in participants’ routine Lighthouse care settings, in either the facilities or CBC settings.

\subsection{Participant Recruitment} 
Eligible PLHIV on ART from the Lighthouse facilities and the CBC settings were identified  at clients' routine care settings by the Lighthouse or CBC study team using purposive sampling ~\cite{tongco2007purposive}. Eligible participants were aged 18 and above, had at least a primary education, and consented to audio recording. In a private location, each eligible participant individually underwent an informed consent process\lrem{, were made aware of Lighthouse’s interest in their responses and experiences, and asked if they understood the study purposes and whether they would like more information}. Appropriate information and explanation of the study goals and objectives were also provided in a consent form written in an accessible language level for improved comprehension. \ladd{The process of obtaining informed consent was prepared by the local staff in Malawi, taking into account the average literacy level of their clients. The consent form highlighted two key points: 1) the voluntary nature of participation in the study, emphasizing that it is entirely up to the individuals to decide whether or not to take part, and 2) ensuring that their decision to participate or not will not have any negative impact on the ART care. The local study staff provided detailed information about the study procedure, including the use of audio recording. The risks and potential benefits of study participation were discussed and questioned answered. After signing the consent form, clients were still able to refuse to answer questions or withdraw from the study. }

Lighthouse is the largest public provider of ART services, covering urban, peri-urban, and rural populations throughout the Lilongwe district. As such, participants were from a diverse client population. In general, 60\% of ART clients in Malawi and at Lighthouse are female. We recruited 63 participants (51 females, 12 males) through Lighthouse’s two urban facilities and four CBC settings. Participants ranged in age from 21 to 73 (mean=47, SD=11.61). Participants were compensated with 10310.00 Malawian Kwacha (approximately \$10 USD) for their participation. We refer to participants from the facilities as P\#-Facility\# and participants from the CBC settings as P\#-CBC\#.

\subsection{Study Procedure}
A total of eight FGDs were conducted: four in facilities and four in CBC settings. Each FGD was composed of 7-9 participants and was facilitated by a Malawian qualitative researcher with support from a notetaker. All FGDs were conducted in Chichewa, the main language used in Malawi. Each FGD began with a short introduction by the facilitator, including the background, purpose, and objectives of the study, and a review of confidentiality and consent guidelines. 

Participants were first asked an ice-breaker question about their preferences for care settings to allow them to gain comfort in speaking. Next, participants were asked about the flow of their appointment visits, what information they shared with providers, and how providers used the \lrem{digital}\ladd{devices for} data collection\lrem{ tools} at their routine care setting (computers at facilities, tablets at CBC settings) during appointments. Participants were also \lrem{questioned}\ladd{asked} to compare the \lrem{digital tools}\ladd{devices} used in their routine care setting to other tools they see being used at Lighthouse, such as paper records (e.g., ``In thinking about how your providers record your information in a computer or on paper, which do you prefer and why?”). Participants were asked about their comfort with their health information being collected with \lrem{digital tools}\ladd{devices}, how they make sense of how the \lrem{digital tools}\ladd{devices} are used (e.g., ``How do you know that the provider is entering your data into the tablets if you cannot see what they are doing?”), and how they think \lrem{digital tools}\ladd{devices} affect care (e.g., ``How do you think the tablets affect the way your providers guide your care?”). 

The next set of questions aimed to understand participants’ mental models of data flow and management. Participants were asked about data storage and transfer (e.g., ``How do you think your health information moves from the computer that the provider uses to other people at Lighthouse?”, ``What do you think happens to the data in the tablet at the end of the day?”). To understand participants’ perceptions towards access to personal health information, participants were further asked to identify entities with whom they would and would not want \lrem{to share data access}\ladd{their data shared with}, explain views on data sharing with the MoH and authorized entities (e.g., ``What do you think the MoH does with the data?”), and share concerns about sensitive information being seen by individuals outside of the health system (e.g., ``How would you feel if your health information is accessed by your family or your neighbors?”). After we informed participants of a Malawian policy that requires local data hosting for local partner digital health implementations, participants were asked about their views on the country’s efforts to protect their data.

Lastly, participants were given time to share feedback for Lighthouse and to ask questions about data security and privacy and data management which were not addressed during the session. Each FGD lasted approximately one hour.

\subsection{Data Analysis}
Audio recordings of FGDs were transcribed in Chichewa and translated into English for analysis. We used a mixture of deductive and inductive thematic analysis ~\cite{braun2006using} to code the transcripts. The first author developed a codebook while conducting a detailed reading of all transcripts and returned to the transcripts to apply the codes. The second author reviewed the code application and resolved disagreements through discussion with the first author. \ladd{The second author did not add codes but helped refine the language and meaning in the initial codes. As a researcher with more than a decade working at and with Lighthouse, the second author also provided a larger view of the environment and history of the responses, helping add nuance and trustworthiness to the codes and their refinements.} Similar codes were grouped into higher level themes, reflecting those suggested a priori from the FGD guide (deductive) and from reviewing participants’ contributions (inductive).\lrem{ Data analysis, including theme formation, sub-themes, and illustrative quotes was reviewed and finalized with the Lighthouse study team during weekly meetings.} \ladd{The themes and their alignment with quotes was discussed at three separate meetings with the Lighthouse study team – fine tuning the themes and illustrative quotes until all believed that the findings and discussion points reflected the diverse perspectives of clients. We tried to tread very carefully away from broad generalities as these opinions and perspectives are limited to these participants – who may not reflect larger norms from this larger ART setting.}

\subsection{Ethical Considerations}
\ladd{The first author worked closely with the in-country team on all aspects of the study design, refining of the study procedure, implementation, and interpretation of the findings. The first author worked closely and on-site with the lead, local qualitative interviewer to refine the FGD guide. The local interviewer is a seasoned qualitative researcher who is part of the Lighthouse team. She previously conducted similar sensitive research among this population. She does not routinely interact with clients but is well acquainted with the community, the clients, the sensitivities, and the clinic setting.}

\section{Findings} 

We organize our findings into three sections. \ladd{We begin with \textit{Disclosure and Privacy} because unintended disclosure of positive HIV status is a major privacy concern for PLHIV. This section}\lrem{\textit{Negotiating Disclosure with Privacy}} describes participants’ \lrem{experiences with }\ladd{views towards} disclosing positive HIV status and \lrem{the privacy considerations they make in the disclosure process}\ladd{the intrinsic relation between privacy and disclosure}. \textit{Confidence in the Health System to Protect Client Privacy} describes participants’ trust in the health system, including healthcare providers and national efforts, to protect client privacy. Lastly, \textit{Perceptions towards Data Collection Tools and Data Management} describes participants’ understanding of computers, tablets, and health passports used during data collection and data management processes.

\subsection{\lrem{Negotiating Disclosure with Privacy}\ladd{Disclosure and Privacy}} 
\ladd{HIV disclosure is a complex challenge and experience that PLHIV face. As such, we first elaborate on participants' perspectives towards disclosure to give context to their concerns around data privacy.}\lrem{In response to questions about who participants are comfortable having their data accessed by, participants reflected on their experiences and decisions around the disclosure of their positive HIV status.}\lrem{ We elaborate on how  disclosure is navigated and PLHIV’s obligation to respect and protect other PLHIV’s disclosure-related decisions}\lremF{.}

\subsubsection{Selective Disclosure} 
\lrem{When considering data privacy, participants understood that known positive HIV status could change the treatment they receive from their communities. Participants were concerned about\textit{``people who cannot keep it to themselves”} and tell others about someone’s positive HIV status (P8-Facility3). }\ladd{When explaining their discomfort with their data being accessed by individuals outside of the health system, participants referred to their fears of unintended disclosure of positive HIV status\lremF{ as explanation}. Participants were concerned about having their positive HIV status known publicly, as there are \textit{``[people] who diminish our respect, they say bad things”} (P4-CBC2). Participants agreed that publicly sharing or exposing personally identifiable information - name, phone number, address - \lrem{publicly }would not only be \textit{``violating one’s human rights”} (P1-CBC1) but also severely impact PLHIV’s well-being, in which \textit{``some may even lose their mind and die as they will be surprised”} (P9-CBC4). The consequences of failing to keep data confidential will extend to the clients' family: \textit{``There is a lot that is being said about us who are HIV positive. Once a person knows that you are positive, they think everyone within your family, children inclusive, are positive. The children are not free, they are called names because the parent is positive” (P4-CBC2).}}

To avoid having their status wrongfully exposed, some participants engaged in selective disclosure, in which they \ladd{\textit{``share the information with the person whom you trust”} (P1-CBC1) and \textit{``consider which people to disclose to as well as the atmosphere if the people will be able to maintain confidentiality”}(P6-Facility2).}\lrem{ carefully chose who to disclose their status to: \textit{``If I had told them that it was HIV, the story could have been different … I consider which people to disclose to as well as the atmosphere if the people will be able to maintain confidentiality”} (P6-Facility2). In general, participants agreed that \textit{``you share the information with the person whom you trust”} (P1-CBC1).} 
\ladd{Disclosing to trusted}\lrem{Trusted} individuals \lrem{can support}\ladd{empowered} clients \ladd{to establish a support system that encouraged them to adhere to treatment. Participants said their friends or family help them} \lrem{by ensuring that they }take their medication, \lrem{getting}\ladd{get} medication refills from Lighthouse, or \lrem{providing}\ladd{provide} transportation money for appointments. \lrem{Participants said that PLHIV should disclose to multiple people in their community because \textit{``[if you] only depend on one person, maybe that person may even die before you”} (P7-CBC4). P}\ladd{Some p}articipants emphasized the importance of disclosing their status to family, especially their children, because \textit{``children need to know how to protect you”} (P4-CBC2) and they can become \textit{``my own doctor”} (P1-CBC4). However, not all PLHIV \lrem{can disclose}\ladd{were comfortable disclosing} to their families\lrem{ as there are varying situations}: \ladd{\textit{``I have young sisters and brothers, but I have never disclosed to them because of how they talk, discriminatory”} (P8-CBC4).}

\lrem{
``In a family where people love each other, it is not difficult … But there are some cases where it is hard to disclose. For [PLHIV], we consider ourselves to be from one family and whenever we hear that one of our friends is not fine, then we need to rush and help, so that when others come, we have already rendered a helping hand.” - P9-CBC4
}

\subsubsection{Disclose to PLHIV for Encouragement and Survival}
\ladd{Participants were not concerned about the consequences of data being shared within the HIV community. In CBC, clients rely on group leader clients for support with assisting when one is sick or retrieving medication. P4-CBC2 explained that they voluntarily \textit{``tell [group leaders] everything, including how you are feeling so that you are on the same page,”} indicating that they share sensitive information with \laddF{other} CBC clients\lremF{ to gain} \laddF{for} medical support.}

\ladd{The HIV community was described as \textit{``one family”} that PLHIV can rely on for unwavering support (P9-CBC4). Disclosure was \lremF{seen}\laddF{viewed} as a critical experience that can \textit{``lead [other PLHIV] to salvation”} (P1-CBC2). As a result, participants wanted to \textit{``counsel those that are keeping [their status] to themselves”} (P8-Facility1). Disclosure of positive HIV status was viewed as a path for creating hope for a healthy future for other PLHIV}\lrem{Disclosure can fulfill a purpose for the self and other PLHIV, including mutual encouragement to seek HIV testing and care as well as creating hope for a healthy future}, as expressed by P2-CBC2:

\begin{quote}
\textit{``I told [my acquaintance], `have you seen how I am? I do take medication. I do not mean that you also take the medication but let's just go so that you can meet the providers.’ We went together to [a clinic] where she tested HIV positive. She was counseled and I also encouraged her. I told her I am healthy, and I started taking medication in 2009. You cannot recognize her today. She is doing fine. She appreciates me whenever we meet.”}
\end{quote}

\lrem{Recognizing that disclosure can \textit{``lead [other PLHIV] to salvation”} (P1-CBC2), participants wanted to \textit{``counsel those that are keeping [their status] to themselves”} (P8-Facility1).}\lrem{At the same time, disclosure}\ladd{Participants also recognized that disclosure} can help one improve\lrem{ their own} health management:

\begin{quote}
\textit{``I was comfortable disclosing to everyone because if I am to keep it to myself, I would have died by now. Because when you disclose to someone about your status, they tend to share with you some other ideas. You also motivate others who come and ask how you managed to open-up, in that case you are helping other people … But when I open-up and shared about my status, I noted that some of them became comfortable and asked me how I take care of myself so that they can do the same.”} - P3-Facility4
\end{quote}

\subsubsection{Obligation to Protect Fellow PLHIV}
\lrem{While discussing client data privacy and disclosure risk, p}\ladd{P}articipants expressed an obligation to protect fellow PLHIV from discrimination and stigma. One participant explained that a specific sticker on a client’s health passport may indicate HIV treatment for an insider who is also on ART: \textit{``if we see that, we do not tell anyone as we were taught [by providers] that we need to maintain confidentiality”} (P4-CBC2). In another anecdote, one participant shared how they encouraged another PLHIV to join Lighthouse’s CBC without revealing that they had \lremF{unintentionally}\laddF{accidentally} come to know of this person’s status:

\begin{quote}
\textit{``I escorted my neighbor to maternity. She gave me the health passport and I happened to be playing with it in my hands. She didn’t tell me that she is positive and on ART, but she saw me with her health passport. I just gave it back … She later opened-up [about her HIV status] when we got home … I told her that I am going to the clinic to get my medication since I am on treatment. That is when she asked me that you are on medication, she said that she is on treatment. She is also one of the CBC members. I had maintained confidentiality at the time and didn’t tell anyone.”} - P5-CBC2
\end{quote}

\subsection{Confidence in the Health System to Protect Client Privacy}
Here, we expand on participants’ perspectives of the health system’s efforts to protect \lrem{patient}\ladd{client} privacy in light of the recognition that client privacy protection is of utmost importance. 

\lrem{\subsubsection{Desire to Have Privacy Protected}}
\lrem{
Participants agreed that sharing personally identifiable information - name, phone number, address - publicly would not only be \textit{``violating one’s human rights”} (P1-CBC1) but would also severely impact people’s well-being, in which \textit{``some may even lose their mind and die as they will be surprised”} (P9-CBC4). In terms of research or the media, information should be shared at a general or aggregate level to \textit{``[make] known the issue that the people are facing and not the person”} (P6-CBC4). One participant stressed the importance of guaranteeing that personally identifiable information will not be wrongfully exposed in research: \textit{``our secret is kept between two people [the client and the researcher], so if there are three people then it is no longer a secret”} (P3-Facility1).
}

\subsubsection{Trust in National Efforts}

\paragraph{Privacy Protection}
After informing participants about the MoH policy to require local data hosting for local partner digital health implementations, participants responded optimistically saying that the policy protects clients, as summarized by P6-Facility2: \textit{``It is good, it is giving us a picture that [the MoH protects] the Malawians and that everyone has the right regarding their health. In addition to that, we are protected.”} P4-CBC2 also believed that the policy helps protect the image of Malawi as a country that is actively protecting client privacy. P7-Facility3 emphasized that Malawians’ information should remain within the country and \textit{``other countries just provide the support”} as opposed to being the main manager of Malawi’s client data, implying how data ownership should be navigated.

At the same time, some participants were uncertain about or misunderstood the intention of the policy. \lrem{One participant}\ladd{P4-Facility3} worried that the policy would hinder data sharing to donors who help fund Malawi’s HIV care\lrem{: \textit{``[Malawi would] not be getting the other [countries’] support as they will assume that there is no HIV in Malawi”} (P4-Facility3)}. Another participant \lrem{worried that privacy protection is not compatible with data sharing:}\ladd{asked,} \textit{``We hear that the drugs that are helping us are from the outside countries, so how do they keep the information confidential so that those outside Malawi wouldn’t know?”} (P2-CBC1)\lrem{. This question implies that the participant is unaware  of the difference between}\ladd{, suggesting that they \lremF{lacked clarity}\laddF{did not understand} that} aggregate data\lrem{, which is shared to other countries,} \ladd{for sharing is deidentified} and client-level data \lrem{, which}is confidential.

\paragraph{Request Assistance for HIV Care Programs}
Participants stressed the importance of sharing data with other countries that assist Malawi’s HIV care program to ensure that the support is \textit{``not one time but throughout”} (P7-Facility3). ART clients rely on the MoH to acquire funding that sustains Malawi’s HIV response (P8-Facility3). Participants expressed concern that the failure to share data and request funds \textit{``may affect the project”} to support HIV care (P7-Facility3) and consequently \textit{``[the clients] may end up being affected”} (P3-CBC4). \ladd{At the same time, participants \lremF{trusted}\laddF{were firm in their belief} that data gathered through research or shared through the media should only be general or aggregate data that \textit{``[makes] known the issue that the people are facing and not the person”} (P6-CBC4).}

\subsubsection{Trust in Lighthouse}
Participants \lrem{heavily placed their trust in the }\ladd{trusted that} Lighthouse staff, especially healthcare providers\lrem{, to} execute their day-to-day work in a manner that respects and protects clients’ privacy. Lighthouse was described as a caring organization that brought comfort to clients: \textit{``Lighthouse is the parent to those of us who have issues. As a parent is expected to provide care to the child at the right time, when we get here, we get what we want easily and properly. We appreciate the support that we get”} (P7-Facility3). Participants  trusted that Lighthouse took precautions to protect client records, as summarized by P7-Facility4: \textit{``This is the headquarters and I noticed that our records are kept safe. They have never given me a wrong file. They always give one their right file. This means our files are well protected and no one can know about another person.”} 

\paragraph{Trust that Healthcare Providers Maintain Confidentiality}
Providers were \lrem{trusted because they were }believed to have good intentions \lrem{and they took an oath }to guard clients’ health and privacy (P7-Facility4). Participants’ confidence in their providers \lrem{stems}\ladd{stemmed} from their trust that providers have never broken confidentiality: \textit{``We have come a long way and have never heard anyone talking about our status, or that the providers disclose that they are coming to give us medication, I have never heard of anything like that. So, I plead that they should continue with that behavior, being respectful”} (P7-CBC1). \ladd{Participants trusted that providers will \textit{``just share with [my family] information but not necessarily regarding my status,”} implying the expectation that private, identifiable data will not be shared (P6-Facility2).}

Participants\ladd{'} \lrem{alluded that their}trust in providers impact\lrem{s}\ladd{ed} client-provider interactions. Some participants \lrem{said that they }avoid\ladd{ed} asking many questions \ladd{during appointments}\lrem{ about their care or treatment} to not appear as if they \lrem{are}\ladd{were} questioning their providers’ authority: \textit{``If we are to ask them questions and show that we are being suspicious then the provider is also a human being and might tell us to go to someone [else] since we don’t trust them”} (P6-Facility2). Providers were \lrem{also }seen to have high status because of their exclusive knowledge of how to use \ladd{devices at the point-of-care}\lrem{the technologies} (P6-CBC4).

\subsection{Perceptions towards Data Collection Tools and Data Management}

Participants explained the purposes and benefits of the computers used at the facilities, tablets used at the CBC settings, and health passports, and shared their understanding of data flow and management in the continuum of care.

\subsubsection{Purposes of \ladd{Devices for}\lrem{Digital} Data Collection\lrem{ Tools}}
\paragraph{Carries Health Information}
Participants recognized that the computers and tablets carry clients’ health information such as when the client tested for HIV, the type of medication administered, when medication is collected, appointment dates, when the client is sick, and lab results that indicate viral load. One CBC participant described how providers enter information into the tablet during the appointment: \textit{``when they see us, they document in the tablet, words like “it’s alright”} (P1-CBC3). \lrem{Although client cannot confirm whether providers are using the tools to enter health information, participants assumed this based on their observations: \textit{``the providers poke the tablet whenever we respond to their questions … about our health”} (P7-CBC4).}

\paragraph{Informs Providers’ Decisions for Care}
Participants agreed that client data in the computers and tablets \lrem{informs providers’ decisions for care}\ladd{aid providers with accomplishing important tasks for care, including calculating adherence to medication, monitoring appointment compliance, and reminding providers to take blood samples}.
Participants assumed that providers \lrem{can access}\ladd{accessed} client history in the computers and tablets, like a \textit{``bank where one can get the bank statement [to] show you your account activities”} (P5-CBC2)\ladd{,}\lrem{. Access to client history was believed to} \ladd{which} help\ladd{s} providers \textit{``follow the right procedure”} (P2-CBC1) and \textit{``provide the right care”} (P7-CBC4), such as \lrem{changing }\ladd{prescribing new} medication\lrem{ after observing in the client records that the current medication is not helping}. \ladd{Outside of appointments, participants believed that the devices were helpful for }\lrem{Participants also thought that the digital tools can inform decisions made outside of the appointments, such as }planning the number of drugs to bring to the CBC settings (P1-CBC1) and transferring data to another provider in case the primary provider is \lrem{not }\ladd{un}available (P7-Facility4).

\paragraph{Uncertainty about \lrem{Tools’}\ladd{Devices'} Purposes}
\ladd{Participants made educated guesses about the usage of devices based on their observations: \textit{``the providers poke the tablet whenever we respond to their questions … about our health”} (P7-CBC4).} \lrem{While participants were able to provide educated guesses about how the digital tools function}\ladd{However}, participants noted that there is little direct communication from the providers \ladd{on how devices are used}. P8-CBC3 said that \textit{``what we know is that the provider documents for us, whether they write what we told them or not, we don’t know,”} suggesting that there is no way for clients to confirm whether providers are correctly recording appointment-related content. P5-CBC3 responded immediately after\lrem{,} saying\ladd{,} \textit{``some of us even wonder what they are doing on the tablet.”} How the data is used after the appointment remains a mystery\lrem{, according to P8-Facility1}: \textit{``after we leave, we do not know what they do\lrem{.}”} \ladd{(P8-Facility1).} 

\subsubsection{Comparing the Benefits of Data Collection Tools} 

\paragraph{\lrem{Digital Tool}\ladd{Device} Benefits: Efficiency}
Participants thought paper records placed a heavier work burden on providers than tablets do: \textit{``It will take [the providers] a lot of time to enter the information on paper unlike a tablet which has a lot of space”} (P3-CBC4). Searching client information with \lrem{digital tools}\ladd{devices} was thought to take minimal effort because participants assumed that all of the important information was stored in them: \textit{``We registered during the first visit. When I reported for the second time, they just asked about my phone number and searched it on their tablet and they were able to find all the information, which shows that it is easy. If it was paper-based, then [my information] could have been lost”} (P7-CBC1).

\paragraph{\lrem{Digital Tool}\ladd{Device} Benefits: Data Safety}
\lrem{The most serious concern that was raised by participants was the risk of losing data recorded on paper because of its susceptibility to damage, while \textit{``information that is kept in a computer stays for a long time. It can’t be deleted”} (P4-Facility3). Data loss was a minor concern with computers and tablets because participants assumed that there were multiple access points to client data.}
Participants agreed that \lrem{digital tools can protect}\ladd{devices protected} data better than paper records. Health passports\ladd{,} which are brought home\ladd{,} are easier to access whereas computer-based documentation is not accessible to most \ladd{because}\lrem{:} \textit{``\lrem{N}\ladd{n}ot everyone who comes [to the facility] knows \ladd{[how to use]} computers.\lrem{ It is not easy for them to know the name of the person or what is documented on the computer}”} (P7-Facility1). Four participants mentioned password protection as a security advantage that computers and tablets have over paper. For example, P1-Facility2 attributed the exclusivity of passwords as a strength of \lrem{digital tools}\ladd{devices}: \textit{``for someone like me who does not know computers, [they] cannot know the password.”} \ladd{The most serious concern with paper was its susceptibility to damage, while \textit{``information that is kept in a computer stays for a long time. It can’t be deleted”} (P4-Facility3). Data loss was a minor concern with computers and tablets because participants assumed that there were multiple access points to client data and there were protection mechanisms in place.}

\paragraph{Health Passport Benefits: Retention of Own Health Records}
The most notable benefit of health passports was that clients maintained access to and control over their health information:
 
\begin{quote}
\textit{``It is our right to access the information. There are others who do not have time to ask the providers about their viral load and that information is kept in the health passport. We can also see the information that has been documented in the health passport … One can lose the health passport, but when you report to the facility, they check and find your [client] ID. This shows that these two approaches are good\lrem{, one}\ladd{. Health passports} keep\lrem{s} the information for themselves, and the other information is kept at the facility \ladd{in the computers}.”} - P8-Facility3
\end{quote}

However, some clinical notes in health passports are difficult to interpret: \textit{``I would have liked it if [our providers] would tell us the diagnosis … Many are the times that we just receive the drugs, but they don’t tell us what the diagnosis is. They even document in italicized format [in the health passports] and not everyone can be able to read”} (P8-CBC3). \lrem{Receiving a layperson’s explanation of their health would encourage clients to}\ladd{Participants wanted a layperson's explanation of their health so that they} \textit{``know the steps that we need to take”} for health management (P7-CBC1).

\subsubsection{Data Management and Data Flow}

\paragraph{Access to Client Data}
\ladd{Participants had a clear notion of who data should and should not be shared with, as PLHIV, providers, and LT program teams are all well aware that inadvertent disclosure has severe consequences for PLHIV's lives.} Participants \lrem{would}\ladd{expressed that they} want their data shared with providers, \ladd{trusted} family members, the MoH, program supervisor and coordinator, district health officers, NGOs, and other countries or donors. \ladd{Participants agreed that} \lrem{Client}\ladd{client} data should not be shared with non-medically trained \lrem{staff}\ladd{personnel} at Lighthouse, such as cleaners, guards, accountants, and other clients. Although these groups cannot access client data, participants were concerned about being seen walking in and out of Lighthouse, which is a widely-known provider of ART services: \textit{``[The cleaners and guards] are talkative\lrem{. Once they are done with their work they sit down and drink thobwa (locally made drink).} \ladd{…} When you are passing by, they will say `don’t admire her hips, she was at Lighthouse’”} (P7-Facility1). \lrem{Participants also thought that there were granular access levels among those with client data access at Lighthouse.} P3-Facility1 explained that \textit{``not everyone is supposed to know your status through the facility”} and that client data should only be shared with staff that \lrem{are}\ladd{is} relevant to data collection. \lrem{P7-Facility4}\ladd{In addition to stressing}\lrem{ stated} the importance of assigning data access levels to job responsibility\lrem{: \textit{``There are those who are supposed to see everything depending on the type of care that they must provide.”} Participants also referred to the hierarchy that determined access levels, but}\ladd{, participants} noted that access should not be \lrem{granted }based solely on position, but \lrem{that individuals should}\ladd{also} \textit{``[having] a specific purpose [to] access the information whilst maintaining confidentiality”} (P6-Facility2). 

\paragraph{Data Consolidation and Transfer}
Through personal experience\ladd{s}, participants made sense of the data flow in the continuum of care. During facility appointments, client\ladd{s} visit \lrem{different}\ladd{multiple} stations to meet different staff and discuss \lrem{different}\ladd{various} information. \lrem{P3-Facility4}\ladd{Participants} observed that their information \lrem{was linked across the computers at different stations : }\textit{``\lrem{When the information gets into the computer, it }gets to everyone who has a computer at Lighthouse\ladd{”}} and can be accessed by entering client ID (P3-Facility4).\lrem{ … Whenever they refer you to another section you tend to find your name in the computer once they click it based on your [client] ID.”} Consequently, information was assumed to be shared between providers\lrem{ too,} so that providers \lrem{\textit{``know the problem that [the client] is facing”} and to}\ladd{can} ensure that \lrem{there is no }important information \lrem{that }is \ladd{not} missing or neglected during appointments (P5-Facility2). 

In terms of storing data collected at CBC settings, although most participants thought the tablets \lrem{accumulated}\ladd{stored} data, participants also recognized that their data was transferred to the facilities’ files or computers: \textit{``One time when I reported to the facility, I would just go and mention my name and the computer would be able to show all the information”} (P8-CBC3). Transferring data to the facilities was considered important for continuing care because \textit{``whenever [\lrem{clients}\ladd{we}] get sick on a day that is not a CBC day, we can report to the facility and find the information in the file”} (P3-CBC2). While only two participants discussed \lrem{the details of how they think }the transfer process\lrem{ occurs}, \lrem{it is worth mentioning that }both participants alluded to the similarity between the tablets and personal mobile \lrem{devices}\ladd{phones}. P6-CBC4 described the way \lrem{Specifically, P6-CBC4 said that the }data is transferred from the tablet to the computer \lrem{in}\ladd{as} \textit{``the way we do with our phones, by forwarding the information … like a message.”}

\section{Discussion}

\ladd{
Our main findings are as follows:
\begin{enumerate}
\item When considering the consequences of data being shared to groups outside of the health system, participants referred to fears associated with unintended disclosure of positive HIV status. PLHIV are deliberate about to whom they disclose their status. Disclosing to other PLHIV, by choice, carried less worry as they trusted one another not to reveal anyone's status.
\item Participants trusted \lremF{in }Malawi's health system to responsibly manage and share sensitive client data. Participants also recognized the vital role Lighthouse plays in protecting their data, noting their expectation that healthcare providers must maintain confidentiality and minimize risks for unwanted disclosure.
\item Participants understood how computers and tablets were used in their care, based on observations they made during appointments and personal experiences using technologies. They expected computers and tablets to be more efficient and secure than paper records, although paper-based health passport booklets are important in that they provide clients with a record of their own health.
\end{enumerate}
}

\ladd{In this section, we}\lrem{ We} discuss \lrem{our findings about }ART clients’ broad privacy views\lrem{. We also} \ladd{and} provide a framework for understanding Lighthouse’s successful mobile device adoption\lremF{, }\laddF{.}\lrem{in light of clients’ positive views towards digital data collection tools.} Lastly, based on our discussion points, we provide recommendations for building a strong digital health system that may be applicable to other LMICs.

\subsection{Broad Privacy Concerns with Living with HIV} 
The health privacy discourse in HCI has highlighted digital security concerns around data management ~\cite{10.1145/2212776.2223812, 10.1145/2531602.2531643, 10.1145/2441776.2441837, 10.1145/1979742.1979918}\lrem{, including how PLHIV navigate }\lremF{\ladd{, while also recognizing the }} \laddF{and the} security and privacy trade-offs \ladd{made} in online spaces ~\cite{10.1145/3290605.3300922, 10.1145/3274450, 10.1145/3313831.3376150, 10.1145/3434171, 10.1145/3415244, 10.1145/2971485.2971542}. Our work contributes a broader view of privacy held by PLHIV \ladd{in the healthcare context and beyond that centers around the influences and impacts of disclosure-related decisions and concerns.} \lrem{that extends beyond digital threats. Our study reveals that participants were more concerned about the consequences of disclosing positive HIV status and the health system’s commitment to protect their data, than the technical dimensions of data security.}

\subsubsection{Disclosure}
\ladd{By having a clear notion of who should and should not access their data, participants implied that they want to control how their data is shared ~\cite{schwartz2015patient, caine2013patients, 10.1093/jamia/ocv014, 10.1093/jamia/ocad058} and more largely, control their disclosure processes. Participants were concerned about social rejection or discrimination that follows inadvertent disclosure of positive HIV status, echoing findings from prior work ~\cite{10.1145/3544548.3581033}. With respect to \textbf{RQ1}, PLHIV took control of their disclosure processes by practicing selective disclosure to minimize the chances of unintended disclosure. At the same time, they expected the health system to uphold confidentiality and manage the sensitive data that clients trust it with.}\lrem{Despite engaging in selective disclosure, there remains the concern that trusted individuals will tell others about one’s positive HIV status. Even} \ladd{However, participants were aware that there were unavoidable risks of unintended disclosure, including being seen visiting}\lrem{at the} Lighthouse facilities\lrem{, where distance from non-PLHIV is maintained, participants feared that their physical presence could breed opportunities for discrimination}. The ever-present risk of inadvertent and unwelcomed disclosure permeates ART decision-making among clients, from choosing where to start care to adhering to treatment.

\lremF{However, }\ladd{\lremF{p}\laddF{P}articipants were not heavily concerned with the privacy risks of disclosing within the HIV community}\lrem{privacy within the HIV community was not a major concern for most participants}. \ladd{Also, with respect to \textbf{RQ1}, PLHIV disclosed comfortably within the HIV community to gain and encourage communal and medical support, and respected other PLHIV's approaches to disclosure. Through the decision to disclose to other PLHIV, participants gained communal support and improved self-management. The inherent trust that bonds the HIV community centers around }\lrem{ PLHIV who disclose to other PLHIV receive communal support and guidance and encourage others to disclose. The understanding that disclosure can benefit the self and others may lie in the inherent trust that bonds the PLHIV community. Similar to what is recorded in prior work }\lrem{\mbox{~\cite{10.1145/3411764.3445313, 10.1145/3415244}}}shared values around confidentiality - to respect each other’s privacy and the expectation that their disclosure and hence status would be contained within the community - \laddF{and} was built into the HIV community\lremF{\ladd{, echoing findings from prior work}}\laddF{, as observed in other contexts} ~\cite{10.1145/3411764.3445313, 10.1145/3415244}. \ladd{By upholding the obligation to protect other PLHIV's status, PLHIV empower their peers to take control of their disclosure processes.}

\subsubsection{Client Trust in the Health System}
Participants wanted their data to demonstrate Malawi’s effective HIV response. \lrem{There was an understanding }\ladd{Participants understood} that their data is used by donors in other countries to assess the overall HIV program and to inform global support levels for Malawi’s ART activities. Although the concerns around data privacy have been seen as a contributor to data sharing hesitancy ~\cite{yeager2014factors, EDEN201644}, participants in our study believed that data privacy and data sharing are compatible with one another, given that there are security assurances. \ladd{With respect to \textbf{RQ2}, PLHIV's interests in contributing their data for HIV program assessment and global ART funding stem from PLHIV's trust in Malawi's health system\lrem{, including the MoH and Lighthouse's healthcare professionals,} to protect client confidentiality.} \lrem{Based on participants’ reliance on healthcare providers to maintain confidentiality, we}\ladd{We} deduce that clients’ trust in data sharing derives from their trust in \ladd{national privacy protection efforts and their} providers\ladd{,} who are the main users of digital and paper tools\ladd{~\cite{esmaeilzadeh2019impacts}}. As clients’ closest point of contact in the health system, providers are responsible for ensuring that their work practices align with client privacy expectations and for enhancing client-provider relations.

\subsection{Successful Mobile Device Adoption in an LMIC Health System}

Our work adds to research that recognizes the viability of mobile devices for healthcare in LMICs and their potential for improving the quality of care despite weaknesses in infrastructure or management ~\cite{10.1145/3411764.3445420, 10.1145/3411764.3445111, hall2014assessing, info:doi/10.2196/mhealth.9671, sadasivam2012development, mccool2022mobile}. \lrem{Compared to attitudes towards technologies used in health in LMICs from over a decade ago ~\cite{cheng2011barriers}, }\ladd{With respect to \textbf{RQ3},}\lrem{participants} \ladd{PLHIV} not only accepted the integration of \lrem{digital tools}\ladd{devices} into care, but also recognized the value of digital \lrem{health to enhance}\ladd{health for enhancing} care delivery\ladd{, though some participants were uncertain about technical security mechanisms and exact usage practices}. Participants understood that digital systems could enable quick access to laboratory results and client outcomes and enhance\lrem{d} \ladd{provider} decision-making, which are features of real-time point-of-care technologies that are shown to contribute to promising results in HIV care ~\cite{tweya2016developing, OLUOCH2012e83, douglas2010using, alotaibi2017impact}. \ladd{These findings stand in contrast to technology acceptability in LMICs from over a decade ago in which handheld devices used for health documentation and surveying were viewed with suspicion and confusion by respondents ~\cite{cheng2011barriers}. Participants' understanding of and acceptance of tablets used in care illustrates an increased familiarity with digital systems in healthcare contexts and mobile devices at large ~\cite{handforth2019digital}. At the same time, some participants were uncertain about how devices were used especially outside of appointments, calling for continued efforts to inform clients of device purpose and security.}

As participants acknowledged, technologies do not function alone: providers and Lighthouse as a whole must continue to prioritize clients' needs in terms of privacy and quality of care. In CBC settings, the expansion of tablet-based care likely stemmed from prior, positive exposure to digital data collection at Lighthouse’s facilities, allowing tablet adoption to be accomplished without creating friction or opposition from clients. With nearly seven years of tablet-based data collection and recent efforts to optimize the process, Lighthouse offers an example of successful integration of mobile devices in healthcare, contrary to many mobile device-based health projects in LMICs that do not venture beyond pilot studies due to the lack of follow-ups and rigorous evaluations ~\cite{hall2014assessing, kemp2018implementation}. 

\subsection{Recommendations for Strong Digital Health Systems in LMICs}
Lighthouse makes data privacy a priority. However, gaps remain and may be greater in contexts or settings where client concerns are not as widely recognized or prioritized. In Malawi and other LMIC contexts, health systems face challenges with privacy regulation and mechanisms for redress. The existing state of health systems and the large population of PLHIV in LMICs call for continued interventions that can improve protocols, policies, training, and supervision. This agenda is further motivated by the nascent digital infrastructure of many LMICs that provides an early opportunity for data security interventions that can be built into health systems, as opposed to being retrofitted. With these considerations in mind and building on participants’ inputs, we provide the following recommendations for building strong digital health systems in LMICs while advocating for continued consideration of clients’ voices.

First, clients should be communicated about their data and how it is used in accessible, non-technical language. Similar to previous research that notes that providing access to data is insufficient if clients are not able to understand their data ~\cite{zhang2020patient, PMID:30094287, 10.4108/icst.pervasivehealth.2013.252109, 10.1145/3154862.3154885}, Lighthouse clients want help to translate health information from their health passports into accessible language. Health information should be presented to clients in layperson’s terms ~\cite{10.1145/2998181.2998357} that enable clients to interpret their data for influencing everyday practices for health management. During clinical encounters, providers should leverage \laddF{devices at} the point-of-care\lremF{ technologies} as ``common information spaces” for interactive conversation between clients and providers, not \ladd{for} one-way documentation of client responses ~\cite{reddy2001coordinating, hughes1997constructing, 10.1145/1978942.1979438}. In terms of data sharing, while some participants understood the difference between aggregate data and client-level data, this distinction was not obvious to all. First and foremost, clients should be educated about aggregate vs. client-level data. To assuage privacy concerns, the health system should provide clients with a transparent disclosure of the way different data is used and educate clients about data protection strategies ~\cite{10.1093/jamia/ocac198}. Specifically, that aggregate data is beneficial for strengthening the health system and informing decision-making, while client-level data benefits client’s individual care when shared across individuals involved with care. At the same time, clients should be ensured that client-level data is never publicly released. We also note that during the FGDs, we informed participants about Malawi’s in-country policy on data hosting which opened up discussion. \lrem{Building on }\ladd{Echoing recommendations from} prior work ~\cite{EDEN201644, esmaeilzadeh2019impacts}, we encourage such \lrem{disclosure}\ladd{explanation} of privacy protection policies and efforts in comprehensible language to build client trust in the health system. 

Second, the health system should leverage clients’ familiarity with technology to enhance the acceptability of \lrem{technologies}\ladd{devices used in care}. As clients are one of the main beneficiaries of \laddF{technologies at the} point-of-care\lremF{ technologies}, their experiences and perspectives should be at the center of technology development and adoption efforts. \lrem{CBC c}\ladd{C}lients had a partial understanding of the role of \lrem{EMRs}\lremF{\ladd{electronic data}}\laddF{digital data}, the purposes of \lrem{digital tools}\ladd{devices} \laddF{at the point-of-care}, and the strengths of \lrem{digital tools}\ladd{devices} over paper records. Overall, because of their familiarity with \ladd{computers \lremF{used }at} \laddF{the} point-of-care \ladd{in facilities}\lrem{ technologies}, CBC clients \ladd{were comfortable with and }accepted\lrem{ the value of} tablets in their care. Based on our findings, we recommend that in addition to considering the digital maturity of an environment - identifying the level of technology that an environment is prepared for, considering the sustainability of technology usage, and setting realistic expectations for technology use ~\cite{FAULKENBERRY2022101111} - the health system should build on clients’ familiarity with technology to ensure that new technologies are accepted by clients. There should be continuous efforts to enhance client familiarity even after technology adoption, by providing clear explanations of the technologies’ purpose and clarifying the technologies’ value ~\cite{knipe2014challenges}. 

Third, technologies should be secure beyond client expectations by meeting global security standards. While some participants had an understanding of various security protocols, such as passwords and access controls, digital security was not \lremF{of}\laddF{a} major concern. Rather, clients were more interested in the health system's efforts to protect client privacy\lremF{ - }\laddF{, namely} policies, execution of data sharing, and provider practices\lremF{ - than the details of digital security} ~\cite{shield2010gradual}. This in turn may reflect an expectation that the health system and its governance will ensure robust, highly-secure\lremF{, health} technologies. Clients’ trust in the health system urges a call to action to ensure that technical security is competently built, assessed, strengthened, and sustained by professionals. Those responsible for eHealth initiatives should ensure that global standards for security\lrem{, such as the Principles for Digital Development ~\cite{DigitalDevelopment} or the Health Data Governance Principles ~\cite{Transform_Health_2022},} are met\lrem{ to enable best practices}. \ladd{These global standards include the Principles for Digital Development ~\cite{DigitalDevelopment} which identify core tenets of addressing security and privacy, and the Health Data Governance Principles ~\cite{Transform_Health_2022}, which outline standards for the usage of data within and across health systems, promote learning from well-established policies and regulations, and outline limitations on data access. }

\subsection{Limitations and Future Work}
While this study has a sufficient sample size of participants to reach data saturation, the findings of this study are not generalizable to all clients at Lighthouse. We also acknowledge that the recommendations we make for building strong digital health systems in LMICs may not necessarily be applicable to all health systems in LMICs, given different levels of maturity with digital health integration. Lastly, we aim to use these findings to support additional data security training for Lighthouse and similar LMICs, providing guidance for improved practices stemming directly from these lessons learned. 

\section{Conclusion}
In LMICs, eHealth initiatives are increasing rapidly \lremF{for}\laddF{because of} their potential to support comprehensive care, including across the HIV care continuum. Focusing on clients’ data privacy perspectives and priorities \lremF{are}\laddF{is} critical for enhancing security and confidence in eHealth \lrem{users and }systems\lremF{ \ladd{and their beneficiaries}}. In an effort to elevate clients’ voices, we explored ART client perspectives on data security and privacy in Malawi’s HIV health system. Our findings suggest that PLHIV have a broad privacy framework that \lrem{is not restricted to technology-related threats, but also encompasses clients’ decisions around status disclosure and their trust in the health system to protect sensitive data}\ladd{centers around concerns of how poorly managed or protected data could lead to unwanted disclosure of positive HIV status. Although disclosure considerations impact various decisions ART clients make, these clients trusted that Lighthouse and Malawi's health system will protect their data and themselves from unintended disclosure. This trust led PLHIV in this context to support digital data collection for client care management, with likely personal benefits, and for data sharing to secure more global health funding -- a benefit for the ART program, overall.}

Overall, this work highlights the importance of building strong digital health systems that can respond to clients’ expectations and trust. \ladd{PLHIV in our study understood the value of devices in their care, although some uncertainties remained. At the same time, technology-related threats to data privacy were not a major concern for PLHIV in this context, demonstrating \lremF{support for}\laddF{their expectation of} the health system\lremF{'s efforts} to build and maintain secure technologies. \lremF{To maintain this trust, h}\laddF{H}ealth systems should continue to make all efforts to ensure that digital health systems are reliably built, assessed, strengthened, and sustained.} Especially in LMICs which are understudied and face resource challenges, ensuring a robust \lrem{and reliable }eHealth system with secure data protections from data collection through sharing of aggregate data is critical for client confidence and participation. We encourage HCI researchers to explore client perspectives in other LMICs to inform decisions for client data protections, develop secure and client-centered technologies, and smoothly integrate eHealth initiatives across data users, from clients to the MoH.

\begin{acks}
Research reported in this publication was partially supported by the National Institutes of Health (NIH), National Institute of Mental Health (NIMH) under award number R21MH127992 (“The Community-based ART REtention and Suppression (CARES) App: an innovation to improve patient monitoring and evaluation data in community-based antiretroviral therapy programs in Lilongwe, Malawi”), under multiple principal investigators, Feldacker and Tweya. The authors would also like to thank the Lighthouse Trust and the MoH, Malawi, for their contributions, insights, and recommendations to this important area of study. The content is solely the responsibility of the authors and does not necessarily represent the official views of the NIH, NIMH. Lighthouse Trust, nor Malawi MoH.
\end{acks}

\bibliographystyle{ACM-Reference-Format}
\bibliography{reference}

\end{document}